\documentclass[aps, prb, onecolumn, superscriptaddress, showpacs, 12pt]{revtex4}
\usepackage{amssymb}
\usepackage{amsmath}
\usepackage{bm}
\usepackage[dvips]{graphicx}
\usepackage{dcolumn}
\usepackage{longtable}
\usepackage{color}
\usepackage[dvipdfm]{hyperref}
\def\vect#1{\mbox{\boldmath $#1$}}
\begin{document}

\title{Robust metastable skyrmions and their triangular-square lattice-structural transition in a high-temperature chiral magnet}
\author{K.~Karube}
\altaffiliation{These authors equally contributed to this work}
\affiliation{RIKEN Center for Emergent Matter Science (CEMS), Wako 351-0198, Japan.}
\author{J.S.~White}
\altaffiliation{These authors equally contributed to this work}
\affiliation{Laboratory for Neutron Scattering and Imaging (LNS), Paul Scherrer Institute (PSI),
CH-5232 Villigen, Switzerland.}
\author{N.~Reynolds}
\affiliation{Laboratory for Neutron Scattering and Imaging (LNS), Paul Scherrer Institute (PSI),
CH-5232 Villigen, Switzerland.}
\affiliation{Laboratory for Quantum Magnetism (LQM), Institute of Physics, \'Ecole Polytechnique F\'ed\'erale de Lausanne (EPFL), CH-1015 Lausanne,
Switzerland.}
\author{J.L.~Gavilano}
\affiliation{Laboratory for Neutron Scattering and Imaging (LNS), Paul Scherrer Institute (PSI),
CH-5232 Villigen, Switzerland.}
\author{H.~Oike}
\author{A.~Kikkawa}
\author{F.~Kagawa}
\affiliation{RIKEN Center for Emergent Matter Science (CEMS), Wako 351-0198, Japan.}
\author{Y.~Tokunaga}
\affiliation{Department of Advanced Materials Science, University of Tokyo, Kashiwa 277-8561, Japan.}
\author{H.M.~R\o nnow}
\affiliation{Laboratory for Quantum Magnetism (LQM), Institute of Physics, \'Ecole Polytechnique F\'ed\'erale de Lausanne (EPFL), CH-1015 Lausanne,
Switzerland.}
\author{Y.~Tokura}
\affiliation{RIKEN Center for Emergent Matter Science (CEMS), Wako 351-0198, Japan.}
\affiliation{Department of Applied Physics, University of Tokyo, Bunkyo-ku 113-8656, Japan.}
\author{Y.~Taguchi}
\affiliation{RIKEN Center for Emergent Matter Science (CEMS), Wako 351-0198, Japan.}

\maketitle

\newpage
\textbf{
Skyrmions, topologically-protected nanometric spin vortices, are being investigated \cite{Bogdanov,Muhlbauer,Yu_FeCoSi,Nagaosa2, Heinze, Romming, Ishiwata, Kezsmarki, Yu_FeGe, Seki, White} extensively in various magnets. 
Among them, many of structurally-chiral cubic magnets host the triangular-lattice skyrmion crystal (SkX) as the thermodynamic equilibrium state. 
However, this state exists only in a narrow temperature and magnetic-field region just below the magnetic transition temperature $T_\mathrm{c}$, while a helical or conical magnetic state prevails at lower temperatures.
Here we describe that for a room-temperature skyrmion material\cite{Tokunaga}, $\beta$-Mn-type Co$_8$Zn$_8$Mn$_4$, a field-cooling via the equilibrium SkX state can suppress the transition to the helical or conical state, instead realizing robust metastable SkX states that survive over a very wide temperature and magnetic-field region, including down to zero temperature and up to the critical magnetic field of the ferromagnetic transition. Furthermore, the lattice form of the metastable SkX is found to undergo reversible transitions between a conventional triangular lattice and a novel square lattice upon varying the temperature and magnetic field. These findings exemplify the topological robustness of the once-created skyrmions, and establish metastable skyrmion phases as a fertile ground for technological applications.
}

\newpage
Non-collinear and non-coplanar spin textures have recently attracted much attention as a source of various emergent phenomena\cite{Tokura,Nagaosa}, such as multiferroicity, topological Hall effect, etc. Among them, magnetic skyrmions are a prototypical example of a non-coplanar spin texture with a nanometer size, emerging from several microscopic mechanisms in various kinds of magnets\cite{Bogdanov,Muhlbauer,Yu_FeCoSi,Nagaosa2, Heinze, Romming, Ishiwata, Kezsmarki,Yu_FeGe,Seki,White}. 
One class of such materials is structurally-chiral magnets, such as MnSi\cite{Muhlbauer}, Fe$_{1-x}$Co$_x$Si\cite{Yu_FeCoSi}, FeGe\cite{Yu_FeGe} with $B$20-type structure and Cu$_2$OSeO$_3$\cite{Seki,White}. These compounds show a long-period helical magnetic state as the ground state, described by a single magnetic modulation vector ($\vect{q}$ vector) whose magnitude is determined mainly by the competition between the ferromagnetic exchange interaction and the Dzyaloshinskii-Moriya (DM) interaction\cite{Dzyaloshinskii,Moriya}. Under magnetic fields, such a chiral magnet undergoes a first-order phase transition to a skyrmion crystal (SkX) state. A single skyrmion is characterized by a topological charge of -1 as defined by $\frac{1}{4\pi}\int\int dxdy \: \vect{\mathrm{n}} \cdot \left(\frac{\partial \vect{\mathrm{n}}}{\partial x}\times\frac{\partial \vect{\mathrm{n}}}{\partial y}\right)$, where $\vect{\mathrm{n}}(x,y)$ is the unit vector along the spin direction. This topological charge describes how many times the $\vect{\mathrm{n}}$ unit vectors for the whole skyrmion wrap the unit sphere, and it vanishes for topologically trivial states, such as helical, conical, and spin-collinear (ferromagnetic) states. The different values of the topological charge indicate that skyrmion and helical (or conical) structures cannot be adiabatically transformed between one another by a continuous deformation process\cite{Nagaosa2}.

Skyrmions are promising for spintronics applications firstly because they are stable due to their topological nature, and secondly because they can be manipulated by an ultra-low current density\cite{Jonietz,Schulz,Yu_Skflow,Iwasaki,Sampaio}.
The recent discovery of skyrmion formation at and above room temperature in a new group of chiral magnets, $\beta$-Mn-type Co-Zn-Mn alloys, has provided a significant step toward applications\cite{Tokunaga}. These materials possess a chiral cubic crystal structure with space group $P$4$_1$32 as shown in Fig. 1(a).  
Below a critical temperature $T_\textrm{c}$ $\sim$ 480 K, Co$_{10}$Zn$_{10}$ displays a helical state with a periodicity of $\lambda$ $\sim$ 185 nm\cite{Tokunaga,Hori}. By partial substitution with Mn, both $T_\textrm{c}$ and $\lambda$ decrease, and the target material of the present study, Co$_8$Zn$_8$Mn$_4$, exhibits $T_\textrm{c}$ $\sim$ 300 K and $\lambda$ $\sim$ 125 nm\cite{Tokunaga}.
Small angle neutron scattering (SANS) and Lorentz transmission electron microscopy (LTEM) measurements\cite{Tokunaga} unambiguously demonstrated the formation of  triangular-lattice SkX (Fig. 2(a)), often described by a triple-$\vect{q}$ coupling of three single-$\vect{q}$ vectors that are rotated by 120$^\circ$ from each other and perpendicular to the magnetic field. 
Similar to the other compounds, the thermodynamical equilibrium SkX state of this compound occupies a narrow region just below $T_\mathrm{c}$ on the $(T, H)$ phase diagram as displayed in Fig. 1(b) (see also Supplementary Fig. S3). Such a limited region of SkX stability is unfavorable from the viewpoint of applications.. However, this issue may be overcome by quenching the equilibrium SkX to lower temperatures and realizing a metastable SkX state by a rapid field-cooling (FC), as was demonstrated by a recent pulse-current controlled transport measurement\cite{Oike} for MnSi which required an extremely fast cooling rate of $\sim$ -100 K/sec. 

In the present paper, we report field-temperature-protocol dependent results of AC magnetic susceptibility and SANS measurements on bulk single crystals of Co$_8$Zn$_8$Mn$_4$. The main results are summarized by the state diagram in Fig. 1(c), showing two important findings: (i) the formation of a highly robust metastable SkX via a FC process, and (ii) a new type of lattice structural transition of the SkX.
The metastable SkX state is realized at low temperatures by FC with a moderate or conventionally slow cooling rate $\sim$ -1 K/min, and starting from the equilibrium SkX state formed at 284 $\leq$ $T$ $\leq$ 300 K, as denoted with the pink arrow in Fig. 1(b). The metastable SkX state survives over a wide $(T, H)$ region down to low temperatures, up to the field-induced spin-collinear (ferromagnetic) phase for positive field, and even down to negative field regions. Furthermore, transformations of the SkX between a conventional triangular lattice and a novel square lattice (Fig. 2(a)) were observed by changing the temperature or magnetic field within the metastable SkX state. 

To demonstrate our findings, first, we summarize in Fig. 2(b) the relation between the real-space magnetic structures and the corresponding SANS patterns. 
In the present SANS measurement, the magnetic field was always applied along the [001] axis of the crystal, and the cryomagnet itself was rotated together with the sample around the vertical axis ($\parallel$ [010]) to realize two configurations: the neutron beam parallel to the [001] axis ($H$ $\parallel$ beam) and perpendicular to the [001] axis ($H$ $\perp$ beam), as also shown at the top of Fig. 2(b). 
Since the preferred orientation of the helical $\vect{q}$ vector in Co-Zn-Mn alloys is along the [100], [010] and [001] axes\cite{Tokunaga}, the (three-domain) helical state and the (two-domain) triangular-lattice SkX state respectively show up as 4 spot and 12 spot patterns in the $H$ $\parallel$ beam geometry. 
A square-lattice SkX state, characterized by double-$\vect{q}$ vectors orthogonal to each other and perpendicular to the magnetic field, also shows up as a 4 spot pattern in the $H$ $\parallel$ beam geometry.
In the $H$ $\perp$ beam geometry, the helical multi-domain state shows 4 spots, the conical state 2 spots on the horizontal axis, and both the triangular and square-lattice SkX states each 2 spots on the vertical axis.

Keeping the above relations in mind, we next consider the results (Fig. 3) of the FC process at 0.04 T, i.e. by way of the thermodynamical equilibrium triangular-lattice SkX region (green region in Fig. 1(b)).
The SANS images in Fig. 3(b) show that the pattern obtained from the equilibrium triangular-lattice SkX generated at 295 K persists down to 200 K. This is a direct demonstration of the realization of the metastable SkX state that exists outside of the equilibrium state for temperatures below 284 K.
The lifetime of this metastable SkX is very long and becomes essentially time-independent below 260 K (Supplementary Fig. S4).
At 120 K, the triangular-lattice SkX pattern has partially transformed into 4 spots. 
At 40 K, the 4 spots become clearer and their $|\vect{q}| (\equiv q)$ values become larger than they were at higher temperatures. 
The 4 spot pattern observed at low temperatures corresponds to a square-lattice SkX. A priori, the 4 spot pattern could be interpreted as evidence for a helical multi-domain state. However, this is ruled out by observing the re-warming process (Fig. 3(c)), where the triangular-lattice SkX is recovered already by 200 K, much below the equilibrium SkX phase ($T$ $\geq$ 284 K). Hence, a transition is indeed observed between a triangular-lattice SkX and a square-lattice SkX, both endowed with finite topological charge, as opposed to a helical multi-domain state with zero topological charge.
In the $H$ $\perp$ beam geometry (Fig. 3(b)), the existence of two horizontal and two vertical spots indicates that there is a coexistence of a conical state (horizontal spots) and the SkX (vertical spots), and the intensity from the conical state decreases as the temperature is lowered.

To examine the SkX structural transition more quantitatively, we show the temperature dependence of the total scattered SANS intensity from the SkX states in Fig. 3(d) (for the details of the analysis, see Supplementary Fig. S2), which clearly demonstrates the transition between the triangular-lattice SkX and the square-lattice SkX to display a large thermal hysteresis. 
As shown in Fig. 3(e), the SkX lattice transformation is accompanied by an enhancement of the characteristic $q$ value, i.e. a reduction of the SkX lattice constant. 
In the high temperature region (150 K $\lesssim$ $T$ $\lesssim$ 295 K), the lattice constant of triangular-lattice SkX is estimated as $a_\mathrm{T}$ = $4\pi/\sqrt{3}*q^{-1}$ $\sim$ 130 nm. In the intermediate temperature region (100 K $\lesssim$ $T$ $\lesssim$ 150 K), the triangular-lattice SkX and the square-lattice SkX coexist with different lattice constants: $a_\mathrm{T}$ $\sim$ 126 nm (120 K) and $a_\mathrm{S}$ = $2\pi *q^{-1}$ $\sim$ 102 nm (120 K), respectively.
In the low temperature region ($T$ $\lesssim$ 100 K), the lattice constant of the square-lattice SkX significantly shrinks and saturates at $a_\mathrm{S}$ $\sim$ 73 nm below 40 K.
The variation of the $q$ value is not limited to the metastable SkX state, but is also observed for the helical state after the zero-field-cooling (ZFC) process, i.e. the helical periodicity $\lambda$  = $2\pi *q^{-1}$ shrinks from 110 nm (290 K) to 72 nm (40 K) (Supplementary Fig. S6). 
As shown in Fig. 3(f), the (triangular and square) SkX state always coexist with the conical state while the intensity of the conical state decreases rapidly below 200 K.

To see how robust the metastable triangular-lattice SkX is, we have further investigated the field dependence of the AC magnetic susceptibility and the SANS intensity at 240 K after FC from 295 K at 0.04 T (see Supplementary Fig. S5 for the details). The AC magnetic susceptibility data clearly exhibit a hysteresis behavior between the initial sweeping and the returning process from the field-induced ferromagnetic phase, demonstrating the metastable nature of the initial state. The triangular-lattice SkX pattern is observed in the SANS data for a wide range of magnetic field up to 0.15 T and down to -0.04 T, in accord with the field region where AC susceptibility shows the hysteresis. In the returning process from the ferromagnetic state where the metastable skyrmions are destroyed, only the conical state shows up in the SANS intensity without any revival trace of the SkX of triangular or square lattice form.

Next, we focus on the field dependence of the metastable SkX lattice form in the low temperature region.
Figure 4(b) shows the SANS images obtained during the field-sweeping processes at 40 K after FC.
Upon field-sweeping to the positive field direction, the 4 spots from the square-lattice SkX lose intensity, accompanied by a reduction of the $q$ value. At the same time ring-like scattering intensity from an (orientationally-disordered) triangular-lattice SkX state appears for $H$ = 0.2 T and 0.28 T, and with the same $q$ value as that in the high temperature region. 
The re-entrance of the triangular-lattice SkX state corresponds to the inflection behavior in the AC magnetic susceptibility as shown in Fig. 4(c) (red triangle). 
On the other hand, in the field-sweeping toward the negative field direction, the intensity from the triangular-lattice SkX state is hardly discernible, and the 4 spot pattern persists down to  $H$ = -0.28 T.
The total scattered intensities and $q$ values are plotted in Fig. 4(d) and Fig. 4(e), respectively. 
The intensities obtained from the 4 spot pattern show a shoulder structure at $H$ $\sim$ -0.18 T (Fig. 4(d)), where a small dip structure is observed in the AC magnetic susceptibility (Fig. 4(c)). These anomalies indicate the direct transition from the square-lattice SkX state to the helical state at $H$ $\sim$ -0.18 T without any intermediate transition to the metastable triangular-lattice SkX state. 
The field-induced triangular-lattice SkX state (as represented by the ring-like scattering intensity) is absent also in the field-sweeping process after ZFC (Supplementary Fig. S8(b)), further indicating that the 4 spot pattern in the FC process is different from the helical multi-domain state with zero topological charge, and should be attributed to the square-lattice SkX with the same topological charge as the triangular-lattice SkX state.
In the returning process from the field-induced ferromagnetic region, only the conical state appears as displayed in Fig. 4(f); namely, all the metastable SkX states irreversibly disappear. 

Having proven the existence of robust metastable SkX states in Co$_8$Zn$_8$Mn$_4$, we now discuss their origin. 
In the case of MnSi, although very rapid cooling is necessary to quench the SkX at ambient pressure\cite{Oike}, a metastable SkX can be realized at a moderate cooling rate under pressure\cite{Ritz}. The critical cooling rate for quenching the high-temperature phase is inversely correlated with the metastability of the quenched state, therefore the metastability of the SkX in MnSi seems to be enhanced under pressure. 
There, local strain caused possibly by slight amounts of pressure inhomogeneity may give rise to the enhanced metastability. 
In the present case, Co$_8$Zn$_8$Mn$_4$ has the randomness of site occupancies, i.e. three different atoms (Co, Zn and Mn) on the two crystallographically distinct sites (8$c$ sites and 12$d$ sites), which may play an important role in realizing the robust metastability. A similar persistence of the SkX down to the lowest temperatures has been observed for Fe$_{1-x}$Co$_x$Si solid-solution alloys\cite{Munzer}, where the origin may be the same as in the present case.

Next, let us consider a plausible origin for the reversible transition between the triangular-lattice SkX and square-lattice SkX. 
The skyrmion lattices have some commonalities with superconducting vortex lattices, in which transitions between triangular and square lattice types have been observed and ascribed to anisotropic nature of the superconducting state\cite{Gilardi,Bianchi}. 
In the present case of the SkX, a plausible trigger for the transition is magnetic anisotropy.
In several theoretical studies\cite{Yi,Park,Lin} it is predicted that easy-plane spin anisotropy or exchange anisotropy can induce the transition from a triangular lattice to a square lattice, and at a specific value of magnetic anisotropy the triangular-lattice SkX is recovered by increasing the magnetic field. 
These  theoretical predictions are in good agreement with the present metastable state diagram of Co$_8$Zn$_8$Mn$_4$ shown in Fig. 1(c) if the magnetic anisotropy increases at low temperatures. 
A non-sinusoidal spin rotation in the square-lattice SkX, as indicated by the higher harmonics of $\vect{q}$ (Supplementary Fig. S2(d)), may be a consequence of such magnetic anisotropy.
In addition, the helical state shows the preferred orientation of the $\vect{q}$ vectors to be along $\left<100\right>$ directions at low temperatures. Therefore, once the metastable SkX state is created and each skyrmion is topologically protected, an increasing magnetic anisotropy as a function of decreasing temperature may drive the observed transformation from the triangular to square SkXs. The full description of this SkX lattice structure transition will need to take into account the particle nature of the individual skyrmion beyond the density-wave picture of the SkX.

The present results have thus experimentally uncovered both a new form of SkX, namely the square lattice, and the possibility that novel phenomena can emerge within metastable SkX states. Such robust metastable SkX states prevailing over a much wider temperature and field region beyond the equilibrium SkX phase region further enhance possibilities for spintronics application of skyrmions.

\section*{Methods}
\subsection*{Sample preparation}
An ingot of Co$_8$Zn$_8$Mn$_4$ composed of several single-crystalline grains was grown in an evacuated quartz tube by Bridgman method, as described in Ref. 12. A part of a single-crystalline grain was cut along the (100), (010) and (001) planes with a rectangular shape of 2.20 mm $\times$ 2.24 mm $\times$ 0.36 mm for the AC magnetic susceptibility measurement, and 4.64 mm $\times$ 3.05 mm $\times$ 1.39 mm for the SANS measurement (Supplementary Fig. S1(a)). 

\subsection*{AC magnetic susceptibility measurement}
AC magnetic susceptibility measurements were performed by a standard mutual-inductance method using homemade Cu coils. The magnetic field was applied along the [100] axis. Due to the difference in shape between the samples used in the AC magnetic susceptibility and the SANS measurements, their relative demagnetization factors are different. In order to eliminate the apparent difference arising from the different demagnetization factors, the field values for the AC magnetic susceptibility measurements are calibrated as $H_\mathrm{c} = 3.7*H$ (see Supplementary Fig. S3).
For all the figures related to AC magnetic susceptibility in the main text and Supplementary, the calibrated value $H_\mathrm{c}$ is used with the notation of $H$.
The scanning condition of temperature and field in the FC process is as follows: (i) degaussing at $T$ = 310 K, (ii) ZFC to $T$ = 295 K, (iii) applying field of $H_\mathrm{c}$ = 0.04 T, (iv) FC to measurement temperatures with the cooling rate of 1.3 K/min $\lesssim$ $-dT/dt$ $\lesssim$ 3.0 K/min across the phase boundary ($T$ = 284 K) between the equilibrium SkX and conical phases. The field-sweeping rate at each stabilized temperature is $d H_\mathrm{c}/dt$ $\sim$ $\pm$0.01 T/min.

\subsection*{SANS measurement}
SANS measurements were performed using the instrument \textit{SANS-I} at the Paul Scherrer Institute, Switzerland. 
The neutron beam was collimated over a length of 18 m before reaching the sample. The scattered neutrons were counted by a 2D position-sensitive multi-detector located at 20 m behind the sample. The neutron wavelength was selected as 10 $\mathrm{\AA}$.
Only for the field-sweeping measurements at 40 K in the $H$ $\perp$ beam geometry, the neutron wavelength was changed to 7 $\mathrm{\AA}$ to obtain a wider coverage of the Bragg scattering in this measurement process. 

The mounted single-crystalline sample was installed into a horizontal field cryomagnet so that the field direction was parallel to the [001] axis. 
Maintaining the $H$ $\parallel$ [001] geometry, the cryomagnet was rotated together with the sample around the vertical axis at the neutron beam line (Supplementary Fig. S1(b)), where the angle between neutron beam and magnetic field is defined as $\omega$.
In each SANS measurement at the configuration of $H$ $\parallel$ beam ($\omega$ = 0$^\circ$) and $H$ $\perp$ beam ($\omega$ = 90$^\circ$), 
$\omega$ was scanned by every 2$^\circ$ step in the range up to $\Delta\omega$ = $\pm$ 20$^\circ$ (so-called ``rocking-scan") in order to collect the reciprocal-space distribution of magnetic Bragg scattering. The observed widths in $\omega$-scanned magnetic Bragg peaks were always broader than $\sim$ 5$^\circ$. Therefore, the SANS images shown in the main text, except for Fig. 3(c), are obtained by summing over the above rocking-scan ranges in order to observe both the intensity and positions of all of the Bragg spots in a single image. 
The scanning condition of temperature and field in the FC process is as follows: (i) degaussing at $T$ = 310 K, (ii) ZFC to $T$ = 295 K, (iii) applying field of $H$ = 0.04 T, (iv) FC to measurement temperatures with the cooling rate of $-dT/dt$ $\sim$ 2.0 K/min across the phase boundary ($T$ = 284 K) between the equilibrium SkX and conical phases. The measurement time at each stabilized temperature and field, including the full rocking-scans in the $H$ $\parallel$ beam and $H$ $\perp$ beam geometries, was approximately 50 min.

\section*{References}

\bibliographystyle{apsrev}


\section*{Acknowledgments}
We are grateful to N. Nagaosa, W. Koshibae, S. Zhang, X. Z. Yu, D. Morikawa, T. Nakajima for fruitful discussions. 
\section*{Author contributions}
Y. Taguchi, H.M.R. and Y. Tokura jointly conceived the project. 
The sample preparation was performed by K.K. and A.K. and Y. Tokunaga.
AC magnetic susceptibility measurements were carried out by K.K., H.O. and F.K.
Small angle neutron scattering measurements were carried out by J.S.W., K.K., N.R. and J.L.G.
The results were discussed and interpreted by all the authors.
\section*{Additional Information}
Supplementary information is available in the online version of the paper. 
Reprints and permissions information is available online at www.nature.com/reprints.
Correspondence and requests for materials should be addressed to K.K.
\section*{Competing financial interests}
The authors declare no competing financial interests.

\newpage

\section*{Figure legends}

\textbf{Figure 1: Crystal structure and state diagrams of Co$_8$Zn$_8$Mn$_4$.} (a) $\beta$-Mn-type crystal structure as viewed along the [111] axis. 
It consists of two crystallographic inequivalent sites, 8$c$ sites (blue circles) and 12$d$ sites (red circles). 
Due to its chirality, an alternative enantiomer ($P$4$_3$32) can be derived from the displayed structure ($P$4$_1$32) through a mirror operation. 
(b) Equilibrium magnetic phase diagram near room temperature in Co$_8$Zn$_8$Mn$_4$ determined by isothermal AC magnetic susceptibility measurements in field-increasing runs after zero-field-cooling (ZFC) from 310 K to the measurement temperatures. The equilibrium skyrmion crystal (SkX) state (green area, see Supplementary Fig. S3(c)) exists between $T_\mathrm{c}$ = 300 K and 284 K, below which the helical or conical states are realized (for the determination of the boundaries, see Supplementary Fig. S7 (c)). The pink arrow represents the measurement path of all the presented field-cooling (FC) measurements starting from $(T, H)$ = (295 K, 0.04 T) in the equilibrium SkX state. (c) Magnetic state diagram of Co$_8$Zn$_8$Mn$_4$ determined by isothermal AC magnetic susceptibility measurements in field-sweeping runs after FC as described with the pink arrow in panel (b). The metastable SkX state (surrounded by blue circle symbols) realized by FC survives up to the boundary with the ferromagnetic region, and even in the negative field region, and consists of a triangular-lattice SkX state (light blue area) and a square-lattice SkX state (pink area), as revealed by small angle neutron scattering (SANS) experiments in this study. For the determination of the boundaries, see Fig. 4(c) and Supplementary Fig. S5(c).

\newpage

\textbf{Figure 2: Chiral magnetic structures in real space and $\vect{q}$ space.} (a) Schematic figures of triangular-lattice SkX (left) and square-lattice SkX (right). 
(b) Schematic figures of magnetic structures in real space in the present sample configuration ($H$ $\parallel$ [001]) and the corresponding SANS patterns in the $H$ $\parallel$ beam and $H$ $\perp$ beam geometries. 
The spot appearing at the center (corresponding to $q_i = 0$) is not shown since it is masked out in the experimental SANS images. 
(i) The helical state forms three types of domains with single-$\vect{q}$ $\parallel$ [100] (blue), [010] (red) or [001] (green), respectively. As a result, 4 spots are observed on the horizontal and vertical axes in both the $H$ $\parallel$ beam and $H$ $\perp$ beam geometries. 
(ii) In the conical state with $\vect{q}$ $\parallel$ $H$ $\parallel$ [001], 2 spots in the $H$ $\perp$ beam geometry are observed on the horizontal axis while no signal is observed in the $H$ $\parallel$ beam geometry. 
(iii) In the ferromagnetic state with $\vect{q}$ = 0, no signal is observed. 
(iv) The triangular-lattice SkX state forms two types of domains with 90$^\circ$ rotation, in which one $\vect{q}$ of the triple-$\vect{q}$ structure is aligned along [100] (blue) or [010] (red), respectively. As a result, 12 spots in the $H$ $\parallel$ beam geometry, and 2 spots on the vertical axis in the $H$ $\perp$ beam geometry, are observed. 
(v) The square-lattice SkX state forms one domain-type with double-$\vect{q}$ $\parallel$ [100] and [010]. As a result, 4 spots on the horizontal and vertical axes in the $H$ $\parallel$ beam geometry, and 2 spots on the vertical axis in the $H$ $\perp$ beam geometry, are observed.

\newpage

\textbf{Figure 3: Temperature dependence of the metastable SkX.} 
(a) Schematic illustration of the measurement process. 
The FC process at 0.04 T and the warming process after FC at 0.04 T are denoted by light blue and pink arrows, respectively.
(b) SANS images in the $H$ $\parallel$ beam and $H$ $\perp$ beam geometries at selected temperatures in the FC process. Note that the full scale of the color plot varies between each panel.
(c) SANS images in the $H$ $\parallel$ beam geometry at selected temperatures in the warming process after FC. Again, the full scale of the color plot varies between panels. In this measurement, the rocking-angle was fixed at $\omega$ = 0$^\circ$ and the temperature was swept continuously.
(d) Temperature dependence of the SANS intensity from the triangular-lattice SkX component (blue triangular symbols), and the square-lattice SkX and helical component (red square symbols) as observed in the $H$ $\parallel$ beam geometry. Data in the FC process and the warming process are denoted by the closed symbols (with solid lines) and the open symbols (with broken lines), respectively.
For the details of the estimation of the intensity, see Supplementary Fig. S2. 
The small increase around 250 K in the red square symbols is due to the tiny helical component, and the increase below 150 K indicates the growth of square-lattice SkX component. The blue triangular symbols below 60 K remain non-zero due to a finite contamination from the broad Bragg spots from the square-lattice SkX. 
(e) Temperature dependence of $q \equiv |\vect{q}|$ of the triangular-lattice SkX component (blue triangular symbols) for $T$ $\geq$ 100 K and the square-lattice SkX component (red square symbols) for $T$ $\leq$ 150 K. For the determination of $q$, see Supplementary Fig. S2.
(f) Temperature dependence of the total scattered SANS intensity measured in the $H$ $\perp$ beam geometry. Purple circle symbols denote the scattering appearing on the vertical axes which is the sum of scattering due to triangular and square-lattice SkXs, and a small fraction of helical order. Yellow diamond symbols denote the scattering appearing on the horizontal axis due to the conical order.

\newpage

\textbf{Figure 4: Field dependence of the metastable SkX at 40 K after FC.} 
(a) Schematic illustration of the measurement process. Field-sweepings from 0.04 T to positive and negative field directions are denoted by pink and light blue arrows, respectively. 
(b) SANS images in the $H$ $\parallel$ beam and $H$ $\perp$ beam geometries at selected fields. 
The full scale of the color plot is fixed for the $H$ $\perp$ beam geometry, but for clarity is varied for the $H$ = 0.28 T data in the $H$ $\parallel$ beam geometry.
(c) Field dependence of AC magnetic susceptibility. 
Field-sweepings from 0.04 T to $\pm$0.5 T (from $\pm$0.5 T to 0.04 T) are shown by red (black) lines.  
The blue triangles show the boundaries of metastable SkX state, and the red triangle shows the boundary between the square-lattice SkX state and the triangular-lattice SkX state. These points are plotted, together with the similar data points at different temperatures below 150 K, in the state diagram in Fig. 1(c). The blue triangle at $H$ = -0.18 T corresponds to a tiny dip structure of the AC magnetic susceptibility, and also to a shoulder-like structure seen in the field-dependence of the SANS intensity shown in panel (d).
(d) Field dependence of the SANS intensities (defined similarly as in Fig. 3(d)) for the triangular-lattice SkX component (blue triangular symbols) and the square-lattice SkX and helical component (red square symbols) in the $H$ $\parallel$ beam geometry. Field-sweepings from 0.04 T to $\pm$0.5 T (from $\pm$0.5 T to 0.04 T) are shown by closed (open) symbols. 
The blue triangular symbols in the range of -0.35 T $\leq$ $H$ $\leq$ 0.1 T remain non-zero due to a finite contamination from the broad Bragg spots of the square-lattice SkX and helical states. 
(e) Field dependence of $q$ (defined similarly as in Fig. 3(e)) of the triangular-lattice SkX component for $H$ $\gtrsim$ 0.15 T (blue triangular symbols) and the square-lattice SkX and helical components (red square symbols). 
(f) Field dependence of the total scattered SANS intensities for the sum of the triangular-lattice SkX, the square-lattice SkX and the helical components (purple circle symbols) appearing on the vertical axis, and the conical component (yellow diamond symbols) appearing on the horizontal axis, in the $H$ $\perp$ beam geometry. Field-sweepings from 0.04 T to $\pm$0.5 T (from $\pm$0.5 T to 0.04 T) are shown by closed (open) symbols. 

\newpage


\begin{figure}[tbp]
\begin{center}
\includegraphics[width=12cm]{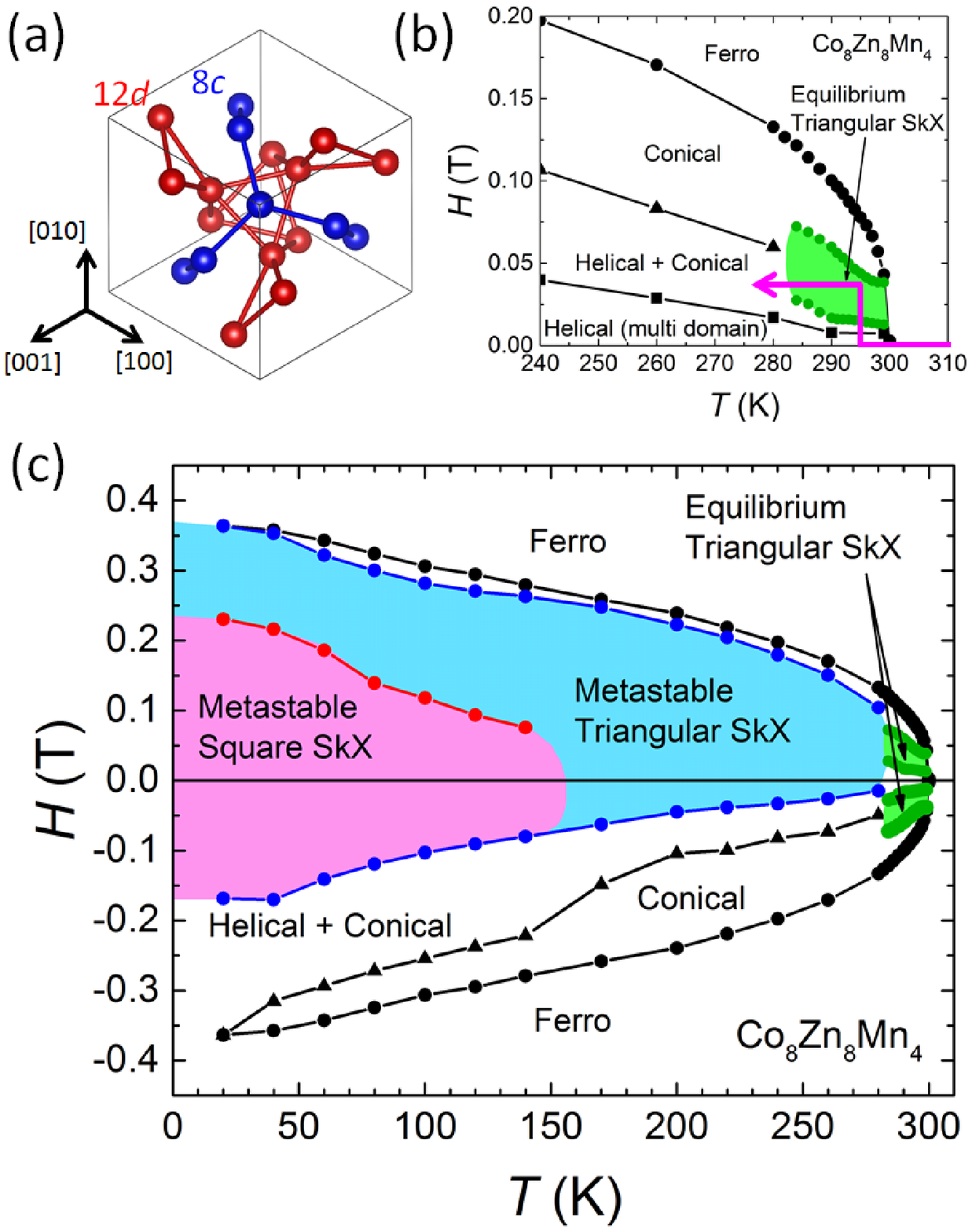}
\end{center}
\end{figure}
\center{\textbf{\Large{Figure 1}}}

\newpage

\begin{figure}[tbp]
\begin{center}
\includegraphics[width=9cm]{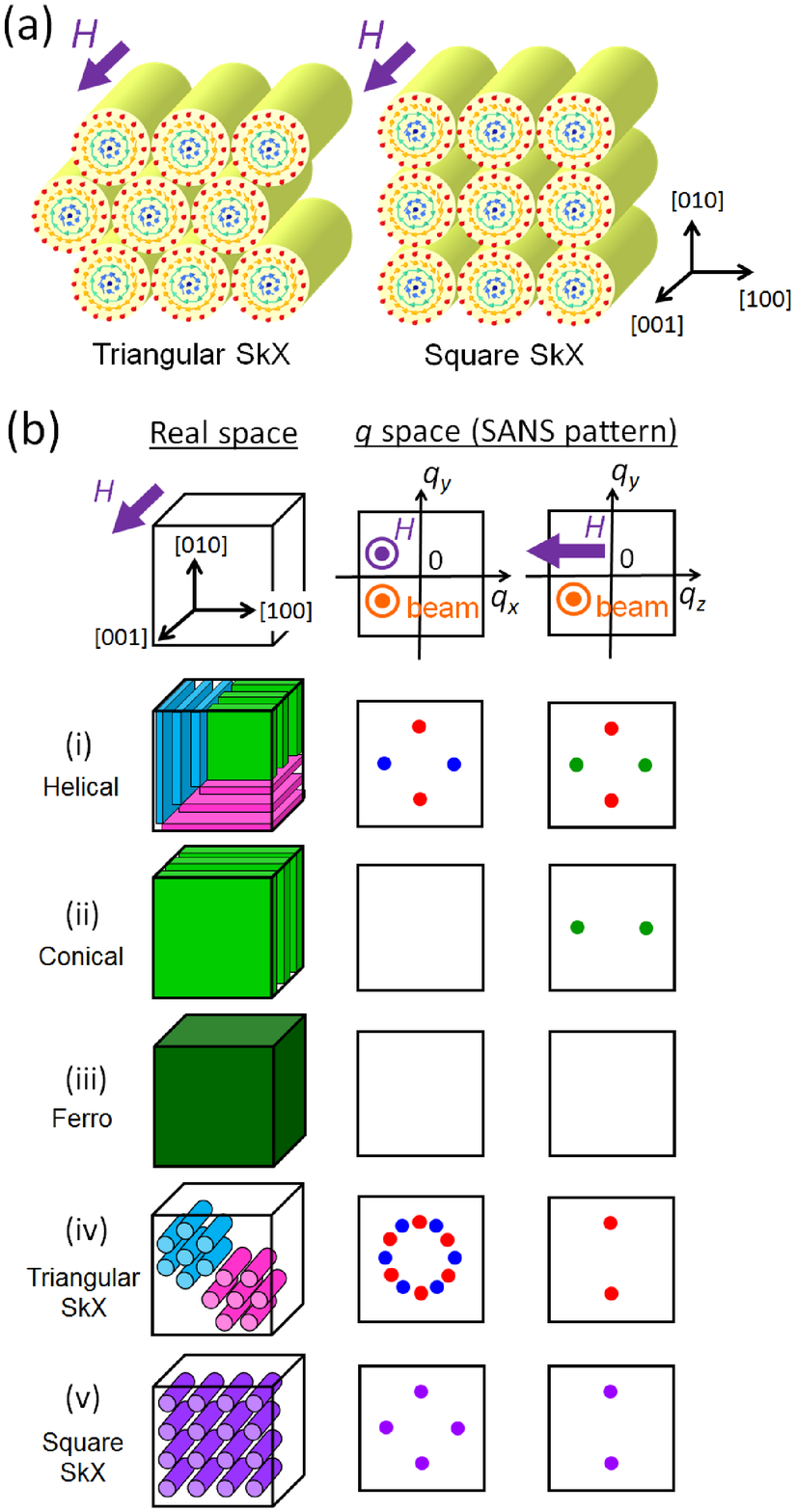}
\end{center}
\end{figure}
\center{\textbf{\Large{Figure 2}}}

\newpage

\begin{figure*}[tbp]
\begin{center}
\includegraphics[width=17cm]{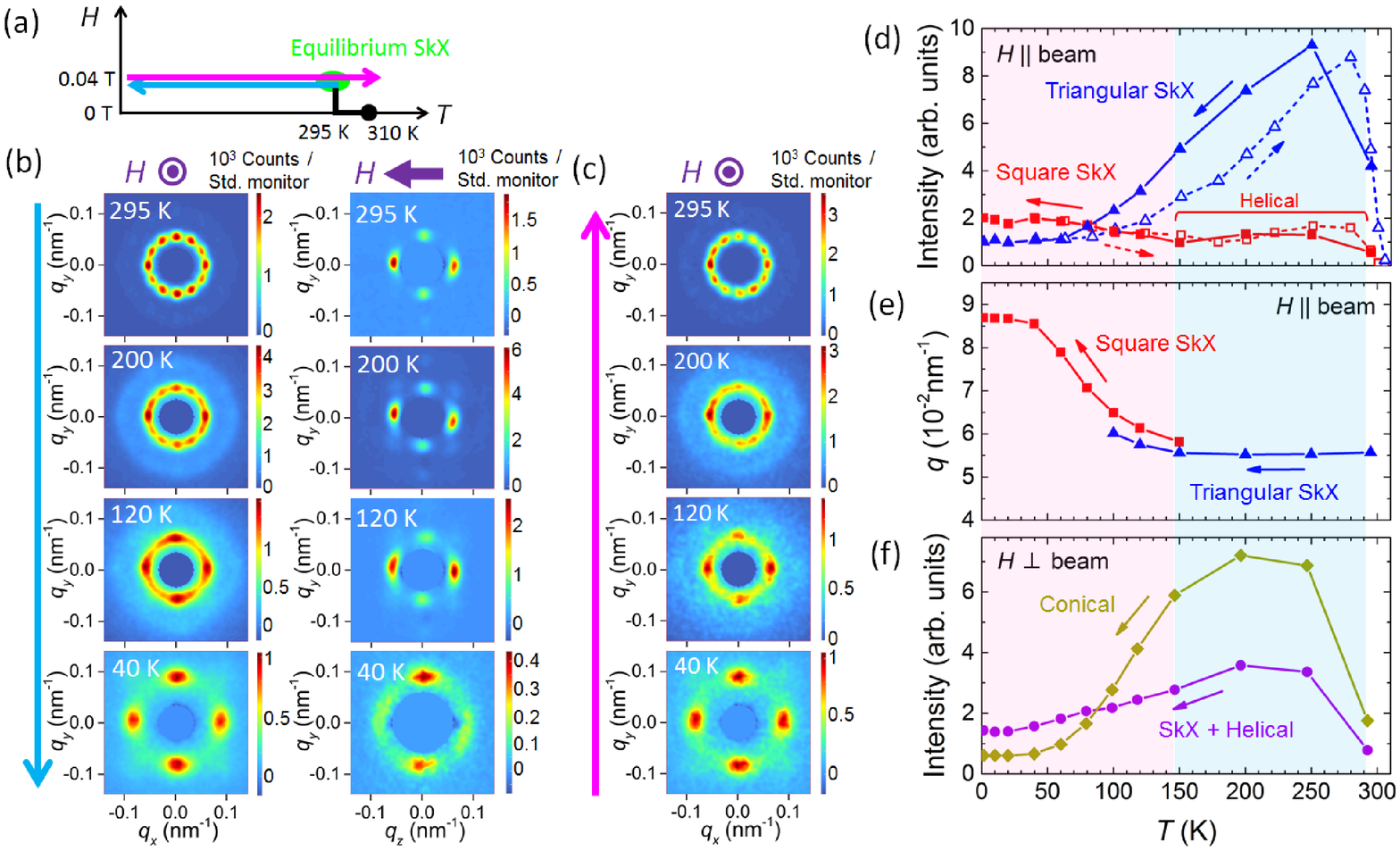}
\end{center}
\end{figure*}
\center{\textbf{\Large{Figure 3}}}

\newpage

\begin{figure*}[tbp]
\begin{center}
\includegraphics[width=16cm]{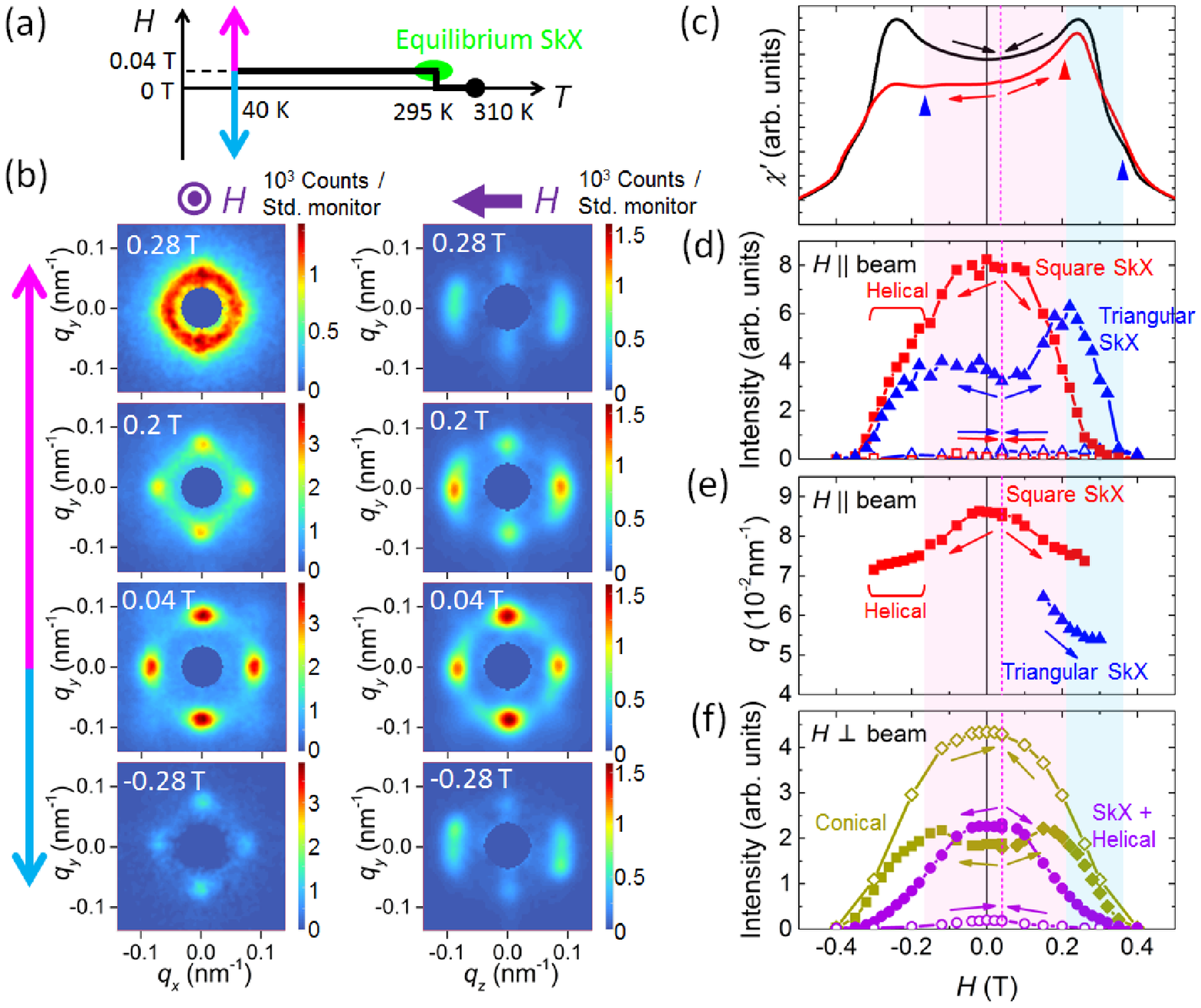}
\end{center}
\end{figure*}
\center{\textbf{\Large{Figure 4}}}

\newpage

\end{document}


\title{Supplementary Information for \\ Robust metastable skyrmions and their triangular-square lattice-structural transition in a high-temperature chiral magnet}
\author{K.~Karube}
\altaffiliation{These authors equally contributed to this work}
\affiliation{RIKEN Center for Emergent Matter Science (CEMS), Wako 351-0198, Japan.}
\author{J.S.~White}
\altaffiliation{These authors equally contributed to this work}
\affiliation{Laboratory for Neutron Scattering and Imaging (LNS), Paul Scherrer Institute (PSI),
CH-5232 Villigen, Switzerland.}
\author{N.~Reynolds}
\affiliation{Laboratory for Neutron Scattering and Imaging (LNS), Paul Scherrer Institute (PSI),
CH-5232 Villigen, Switzerland.}
\affiliation{Laboratory for Quantum Magnetism (LQM), Institute of Physics, \'Ecole Polytechnique F\'ed\'erale de Lausanne (EPFL), CH-1015 Lausanne,
Switzerland.}
\author{J.L.~Gavilano}
\affiliation{Laboratory for Neutron Scattering and Imaging (LNS), Paul Scherrer Institute (PSI),
CH-5232 Villigen, Switzerland.}
\author{H.~Oike}
\author{A.~Kikkawa}
\author{F.~Kagawa}
\affiliation{RIKEN Center for Emergent Matter Science (CEMS), Wako 351-0198, Japan.}
\author{Y.~Tokunaga}
\affiliation{Department of Advanced Materials Science, University of Tokyo, Kashiwa 277-8561, Japan.}
\author{H.M.~R\o nnow}
\affiliation{Laboratory for Quantum Magnetism (LQM), Institute of Physics, \'Ecole Polytechnique F\'ed\'erale de Lausanne (EPFL), CH-1015 Lausanne,
Switzerland.}
\author{Y.~Tokura}
\affiliation{RIKEN Center for Emergent Matter Science (CEMS), Wako 351-0198, Japan.}
\affiliation{Department of Applied Physics, University of Tokyo, Bunkyo-ku 113-8656, Japan.}
\author{Y.~Taguchi}
\affiliation{RIKEN Center for Emergent Matter Science (CEMS), Wako 351-0198, Japan.}
\maketitle

\newpage
\section*{Sample setup}
We prepared single-crystalline samples of Co$_8$Zn$_8$Mn$_4$ for the AC magnetic susceptibility and SANS measurements as shown in Fig. S1(a).
These samples with plate-like shapes were cut out from the same batch grown by the Bridgman method.
In the AC magnetic susceptibility measurement, the magnetic field was applied along the [100] axis, parallel to the plate. On the other hand, in the SANS measurement, the magnetic field was applied along the [001] axis, perpendicular to the plate. 

In the SANS measurement, the cryomagnet was rotated together with the sample around the vertical axis ($\parallel$ [010]) at the neutron beam line, as shown in Fig. S1(b), where the angle between neutron beam and magnetic field is defined as $\omega$.
In each SANS measurement with the configuration of $H$ $\parallel$ beam ($\omega$ = 0$^\circ$) and $H$ $\perp$ beam ($\omega$ = 90$^\circ$), 
the rocking-scan was performed, i.e. $\omega$ was scanned in 2$^\circ$ step over the range $\Delta\omega$ = $\pm$ 20$^\circ$.

\newpage
\section*{Analysis of SANS data}
In many SANS images measured in the $H$ $\parallel$ beam geometry, a 12 spot pattern from the triangular-lattice SkX state (left panel of Fig. S2(a)) and a 4 spot pattern from the square-lattice SkX state or helical state (right panel of Fig. S2(a)) overlap, as seen in Fig. S2(b), which shows representative data taken at 120 K during the FC process. 
We estimated the intensity and $q$-value for each component from the SANS data summed over all rocking angles, and by the follow procedure.

First, the scattered intensity as a function of $q$, and averaged over an azimuthal angle range, $I(q, \theta_0) = \int_{\theta_0-\Delta\theta/2}^{\theta_0+\Delta\theta/2} I(q, \theta)d\theta/\Delta\theta$, is estimated at $\theta_0 = 30^\circ$ (indicated by a blue frame in Fig. S2(b)) and $\theta_0 = 0^\circ$ (indicated by a purple frame in Fig. S2(b)) with $\Delta\theta$ = 10$^\circ$. 
Here, $\theta$ is defined as the azimuthal angle (clockwise) from the $q_y$ axis in the $H$ $\parallel$ beam geometry.
$I(q, \theta_0 = 30^\circ)$ consists mostly of the triangular-lattice SkX component, but at lower temperatures it may also have a small contamination due to the big spot size of the square SkX. Such a contamination can be inferred simply by looking at Fig. S2(b).
On the other hand, $I(q, \theta_0 = 0^\circ)$ consists of triangular-lattice SkX, square-lattice SkX, and helical components. 
Due to the symmetric intensity distribution of the 12 spots intensities of the triangular-lattice SkX, the operation $I(q, \theta_0 = 0^\circ) - I(q, \theta_0 = 30^\circ)$ leaves the intensity that can be attributed mainly to the square-lattice SkX and helical components.

Second, $I(q, \theta_0 = 30^\circ)$ and $I(q, \theta_0 = 0^\circ) - I(q, \theta_0 = 30^\circ)$ are fitted to two Gaussian functions plus a background, as shown in Fig. S2(c) and Fig. S2(d), respectively. The Gaussian area and center in the main peak are defined as $A_1$ and $q_1$, and those in the second peak are defined as $A_2$ and $q_2$, respectively. These $A_1$ and $q_1$ are the total scattered intensity and $q$ which are plotted against temperature or field in Figs. 3, 4 and S5-8. 
The second peak in the triangular-lattice SkX component with the relation of $q_2$ $\sim$ $\sqrt{3} q_1$ (Fig. S2(c)) derives from the vector sum of two of the triple-$\vect{q}$ reciprocal lattice basis vectors from the other domain. 
The second peak in the square-lattice SkX component with the relation of $q_2$ $\sim$ $2 q_1$ (Fig. S2(d)) arises from the higher harmonics of $q_1$, indicating the deviation of the spin modulation from a simple sinusoidal function. 
\section*{Identification of equilibrium SkX state}
We identified the equilibrium SkX state by AC magnetic susceptibility and SANS measurements in the field-sweeping process at 290 K, as shown in Fig. S3. At 290 K after ZFC from 310 K, the helical state appears, showing the 4 spot pattern in Fig. S3(b). With increasing field, the triangular-lattice SkX state with 12 spots appears in the range of 0.03 T $\leq$ $H$ $\leq$ 0.07 T, where the AC magnetic susceptibility shows a dip structure as displayed in Fig. S3(c) with a calibrated magnetic field axis. In the returning process from the field-induced ferromagnetic state at 0.12 T, the triangular-lattice SkX state appears in the slightly lower field range of 0.02 T $\leq$ $H$ $\leq$ 0.06 T than in the field-increasing run. The total scattered SANS intensity estimated to arise from the triangular-lattice SkX is plotted against field in Fig. S3(d). 
\section*{Long lifetime of metastable SkX}
Here we estimate quantitatively the lifetime of the metastable SkX state. 
Figure S4 shows the time dependence of the normalized AC magnetic susceptibility, defined as $R(t) \equiv (\chi_\infty^\prime - \chi^\prime (t))/(\chi_\infty^\prime - \chi_0^\prime)$, after FC down to 282 K and 260 K. 
Here, $\chi_0^\prime$ and $\chi_\infty^\prime$ are the initial value (metastable SkX state) and a fully relaxed value (estimated as the value of the equilibrium helical or conical state), respectively. 
The data points are fitted to a stretched exponential function, $\exp \left\{ -(t/\tau)^\beta \right\}$.
Even at 282 K, just below the boundary of the equilibrium SkX state, the relaxation time (the lifetime of metastable SkX state) from the metastable SkX state to the equilibrium helical or conical state is quite long, estimated as $\tau$ $\sim$ $10^8$ sec. Since the lifetime is inversely correlated with the critical cooling rate for quenching the high-temperature phase (Ref. S1), the moderate cooling rate ($dT/dt$ $\sim$ -1 K/min) is sufficient to quench SkX in the present case. At 260 K, the relaxation behavior is hardly observable due to the much longer relaxation time than the measurement time. This is because the relaxation time increases in an exponential manner as the temperature is lowered (Arrhenius law).
Therefore, in the present SANS measurements performed at stabilized temperatures below 250 K, the time variation of the metastable SkX state is negligible.

\section*{Robust metastable SkX}
The field dependence of the AC magnetic susceptibility and the SANS intensity was investigated at 240 K after FC from 295 K at 0.04 T (Fig. S5). 
Figure S5(c) shows the AC magnetic susceptibility data, clearly displaying a hysteresis behavior between the initial sweeping and the returning sweeping from the field-induced ferromagnetic phase ($\pm$0.3 T). This indicates the metastable nature of the initial state in the blue-shaded region. 
Figure S5(b) shows the field-variation of the associated SANS patterns. In the field-sweeping toward the positive field direction, the metastable triangular-lattice SkX state with the 12 spot pattern survives up to 0.15 T, just before entering the field-induced ferromagnetic state. In the field-sweeping toward the negative field direction, the helical state with the 4 spot pattern gains intensity but the intensity from the triangular-lattice SkX is still detected, even down to $H$ = -0.04 T, as clearly exemplified by the total scattered SANS intensities plotted in Fig. S5(d). The intensity distribution for $H$ $\parallel$ beam and $H$ $\perp$ beam geometries is different for $H$ = $\pm$0.15 T. During these field-sweeping measurements, the $q$ values do not change much, as shown in Fig. S5(e). 
The triangular-lattice SkX is absent in the field-sweeping process after ZFC avoiding the equilibrium SkX state (see Supplementary Fig. S7). 
In the returning process from the ferromagnetic region where the metastable skyrmions are completely destroyed, SANS intensity is only observed from the conical state as displayed in Fig. S5(f). 
These SANS results are in accord with the AC magnetic susceptibility in terms of both their asymmetry with respect to the sign of magnetic field, and their history-dependence. 

\section*{Helical state in the ZFC process}
Figure S6 shows a ZFC process from 310 K. The helical state with 4 spot SANS pattern shows almost the same $q$-value and temperature dependence as those patterns observed from the square-lattice SkX state in the FC process (Fig. S6(d)). 
Figure S7 shows the field-sweeping process at 240 K after ZFC. The helical state persists up to 0.08 T after which it transforms into the conical state. The field dependence of the AC magnetic susceptibility (Fig. S7(c)) and the SANS intensity (Fig. S7(d)) is totally different between the ZFC and the FC processes, while the respective $q$-values remain similar. 
Figure S8 shows the field-sweeping process at 40 K after ZFC. Here the helical state persists up to 0.35 T just below the field-induced ferromagnetic state. The $q$-value and its field dependence in the helical state are also very close to those displayed by the square-lattice SkX state (Fig. S8(e)). 
However, the 12 spot pattern (or ring-like pattern) from the triangular-lattice SkX state as displayed in Fig. 4(b) was never observed around 0.3 T in this process, indicating that the state from which the 4 spot pattern is observed here is distinct from that observed in the FC process, namely, the metastable SkX state.

\section*{References}
S1. Oike, H. \textit{et al}. Interplay between topological and thermodynamic stability in a metastable magnetic skyrmion lattice. \textit{Nat. Phys.} \textbf{12}, 62-67 (2016).

\newpage

\begin{figure}[tbp]
\begin{center}
\includegraphics[width=15cm]{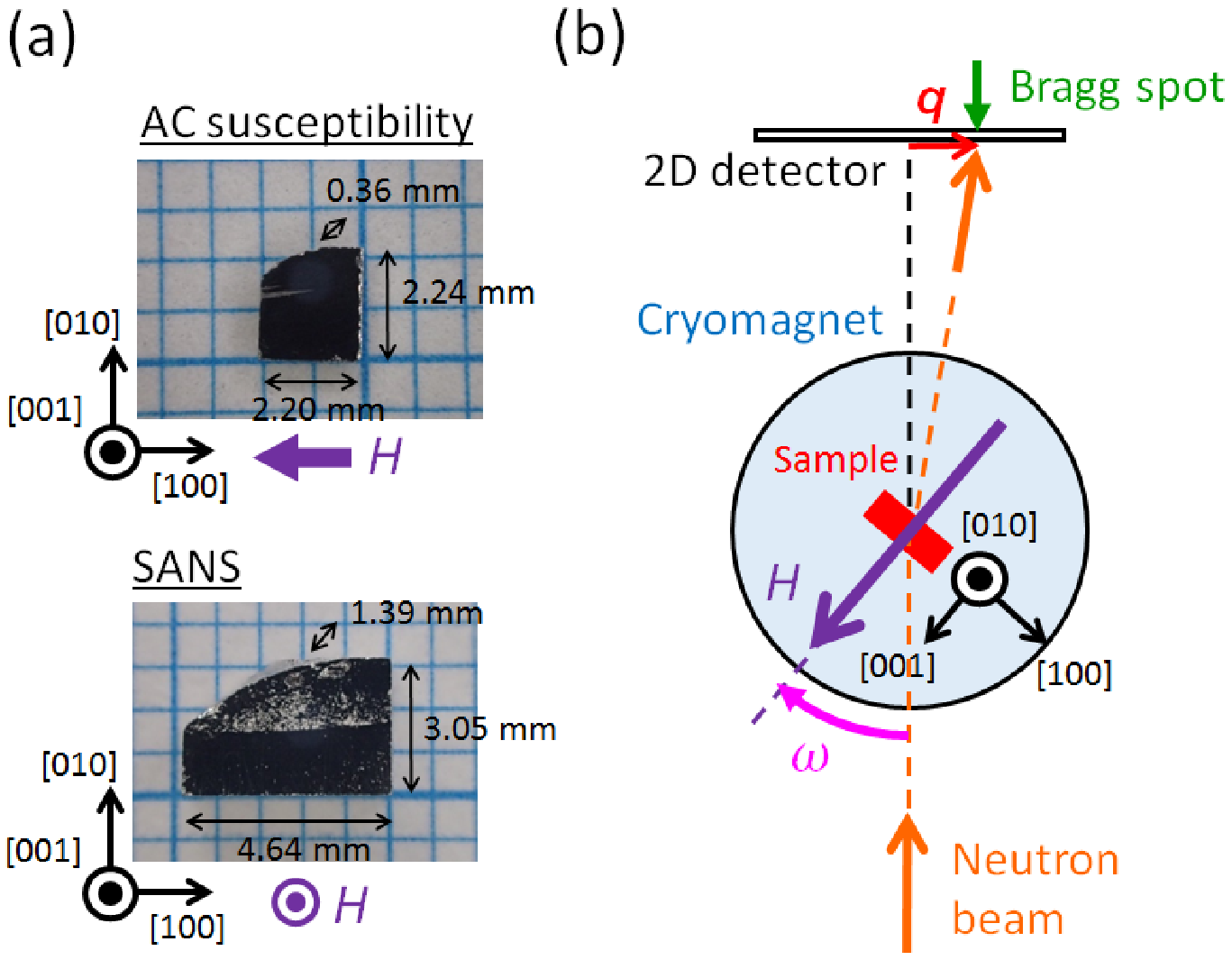}
\end{center}
\end{figure}
\textbf{Figure S1: Sample and SANS measurement configuration.} (a) Pictures of single-crystalline samples of Co$_8$Zn$_8$Mn$_4$. Directions of the magnetic field applied in AC magnetic susceptibility (top) and SANS (bottom) measurements are also indicated. 
(b) Schematic top-view illustration of the SANS measurement configuration. 
\newpage

\begin{figure}[tbp]
\begin{center}
\includegraphics[width=17cm]{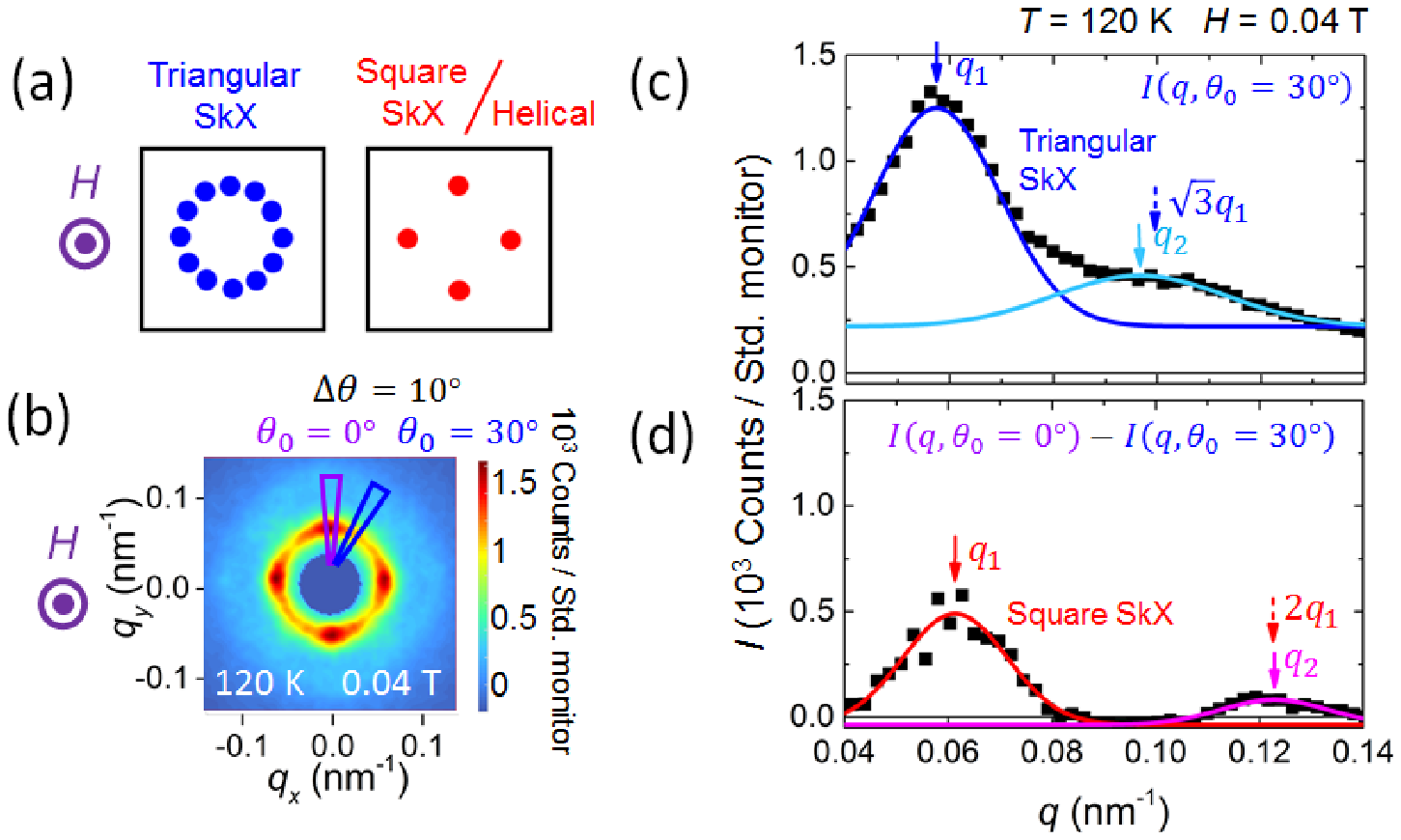}
\end{center}
\end{figure}
\textbf{Figure S2: Analysis of SANS data} (a) Schematic figures of the 12 spot SANS pattern from (multi-domain) triangular-lattice SkX state, and the 4 spot SANS pattern from a square-lattice SkX state or (multi-domain) helical state in the $H$ $\parallel$ beam geometry. (b) SANS image at $T$ = 120 K in the FC process in the $H$ $\parallel$ beam geometry as representative data showing the coexistence of the triangular-lattice SkX and square-lattice SkX states. 
We define the azimuthal angle (clockwise) from the positive $q_y$ axis as $\theta$.
The intensity as a function of $q$ averaged over an azimuthal angle range, given by $I(q, \theta_0) = \int_{\theta_0-\Delta\theta/2}^{\theta_0+\Delta\theta/2} I(q, \theta)d\theta/\Delta\theta$, is estimated at $\theta_0 = 30^\circ$ (blue frame) and $\theta_0 = 0^\circ$ (purple frame) with $\Delta\theta = 10^\circ$. 
(c) $I(q, \theta_0 = 30^\circ)$ at $T$ = 120 K in the FC process (black square symbols) and two Gaussian fitting curves (blue and light blue lines) for the triangular-lattice SkX component with a background. 
(d) $I(q, \theta_0 = 0^\circ) - I(q, \theta_0 = 30^\circ)$ at $T$ = 120 K in the FC process (black square symbols) and two Gaussian fitting curves (red and pink lines) for the square-lattice SkX component. 

\newpage

\begin{figure}[tbp]
\begin{center}
\includegraphics[width=17cm]{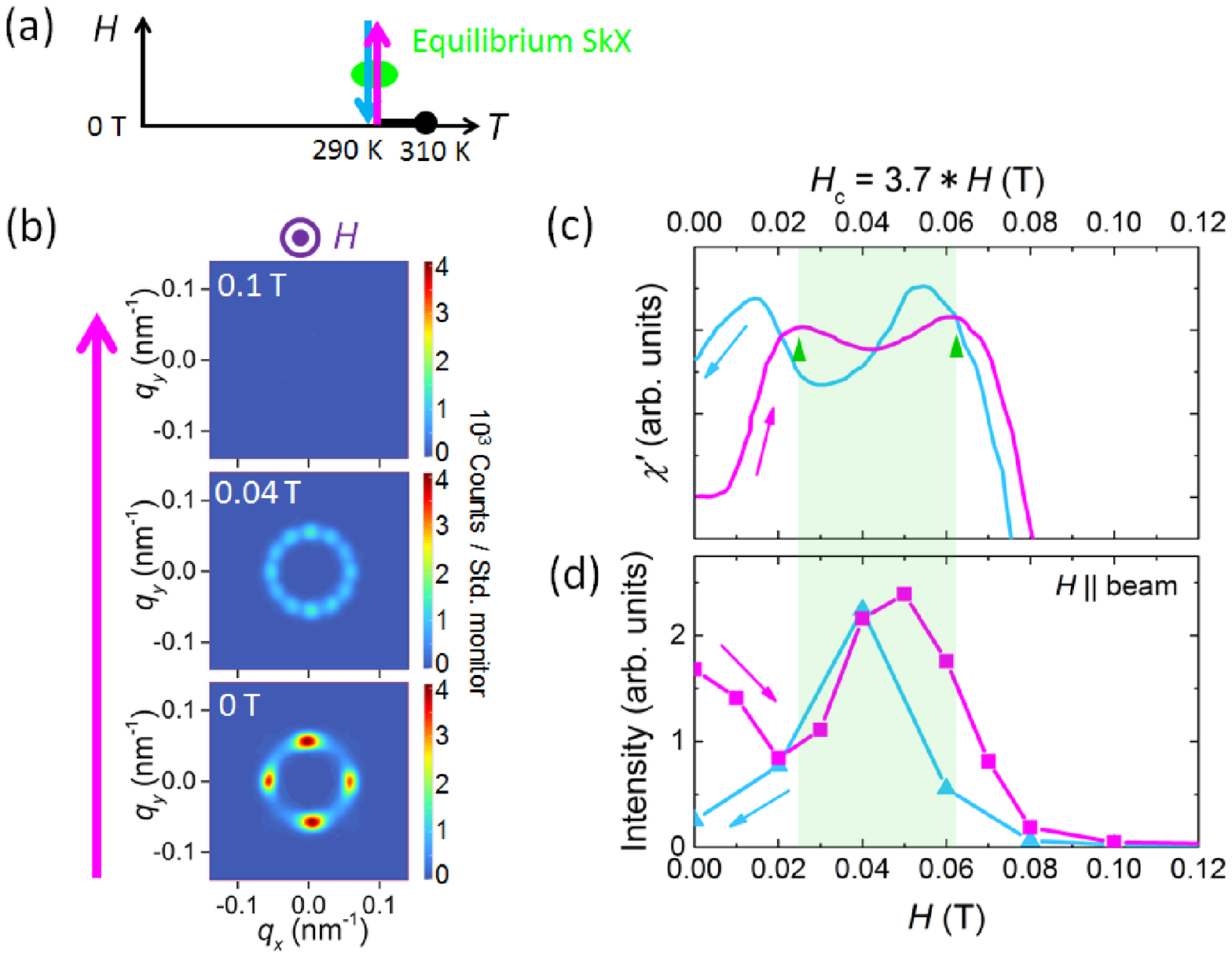}
\end{center}
\end{figure}
\textbf{Figure S3: SANS and AC magnetic susceptibility results in the field-sweeping process at 290 K after ZFC} 
(a) Schematic illustration of the measurement process. Field-sweeping from 0 T to 0.12 T (from 0.12 T to 0 T) is denoted by the pink (light blue) arrow.
(b) SANS images obtained in the $H$ $\parallel$ beam geometry and at selected fields. 
The full scale of the color plot is fixed.
(c) Field dependence of the AC magnetic susceptibility. 
Field-sweeping from 0 T to 0.12 T (from 0.12 T to 0 T) is shown by the pink (light blue) line. 
The green triangles show the boundaries of the equilibrium triangular-lattice SkX state as plotted in the state diagrams shown in Figs. 1(b) and (c). 
Since the demagnetization factor is different from that in the SANS measurement, the field values are calibrated as $H_\mathrm{c} = 3.7*H$.
(d) Field dependence of the total scattered SANS intensity from the triangular-lattice SkX state in the $H$ $\parallel$ beam geometry (defined similarly as in Fig. S2(c)). Field-sweeping from 0 T to 0.12 T (from 0.12 T to 0 T) is shown by pink square (light blue triangle) symbols. 
The pink square symbols for 0 $\leq$ $H$ $\leq$ 0.02 T are non-zero due to a finite contamination from the broad Bragg spot from the helical state. 
\newpage

\begin{figure}[tbp]
\begin{center}
\includegraphics[width=12cm]{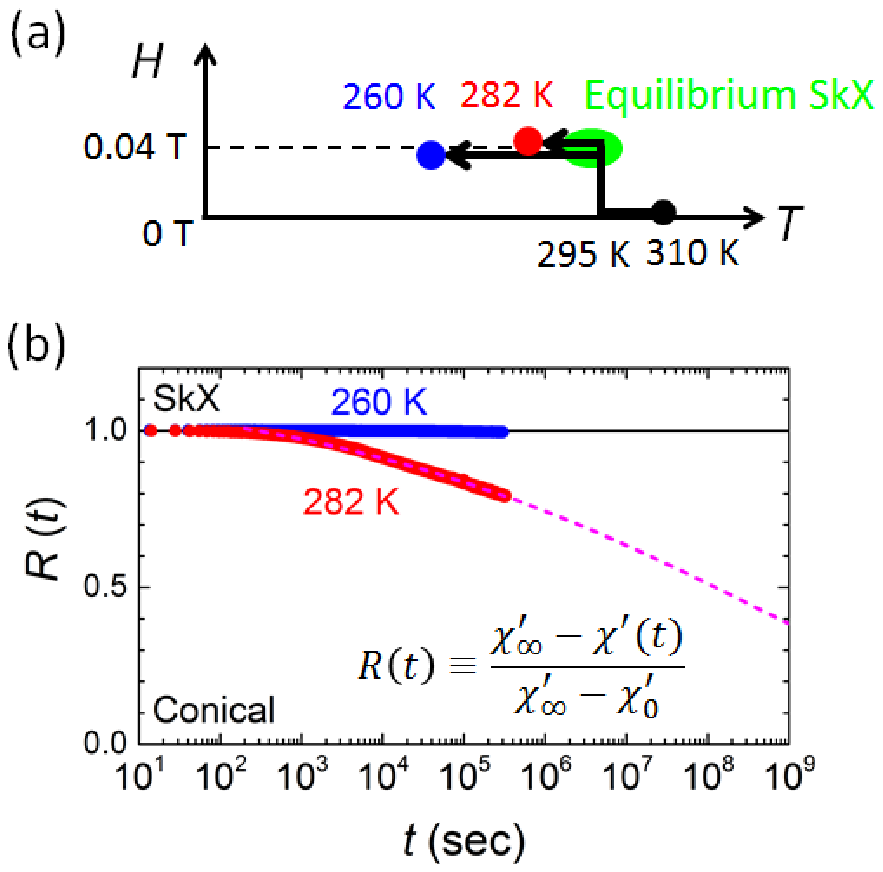}
\end{center}
\end{figure}
\textbf{Figure S4: Time dependence of AC magnetic susceptibility after FC} 
(a) Schematic illustration of the measurement processes. 
(b) Time dependence of the normalized AC magnetic susceptibility, after FC at 0.04 T, at 282 K (red circle symbols) and 260 K (blue circle symbols). The vertical axis is defined as $R(t) \equiv (\chi_\infty^\prime - \chi^\prime (t))/(\chi_\infty^\prime - \chi_0^\prime)$, where $\chi_0^\prime$ ($t = 0$) and $\chi_\infty^\prime$ ($t \rightarrow \infty$) are an initial value [metastable SkX state ($R$ = 1)] and a fully relaxed value [equilibrium helical or conical state ($R$ = 0)], respectively. 
Since the relaxation time is much longer than the measurement time, $\chi^{\prime}_\infty$ is assumed to be the value of $\chi^{\prime}$ at $H$ = 0.04 T in the field-sweeping run from the field-induced ferromagnetic region.
The data points at 282 K are fitted to a stretched exponential function, $\exp \left\{ -(t/\tau)^\beta \right\}$, yielding $\tau = 4.4 \times 10^8$ sec and $\beta$ = 0.13 (broken pink line). The obtained $\beta$-value indicates the existence of a distribution of relaxation times.

\newpage

\begin{figure}[tbp]
\begin{center}
\includegraphics[width=15cm]{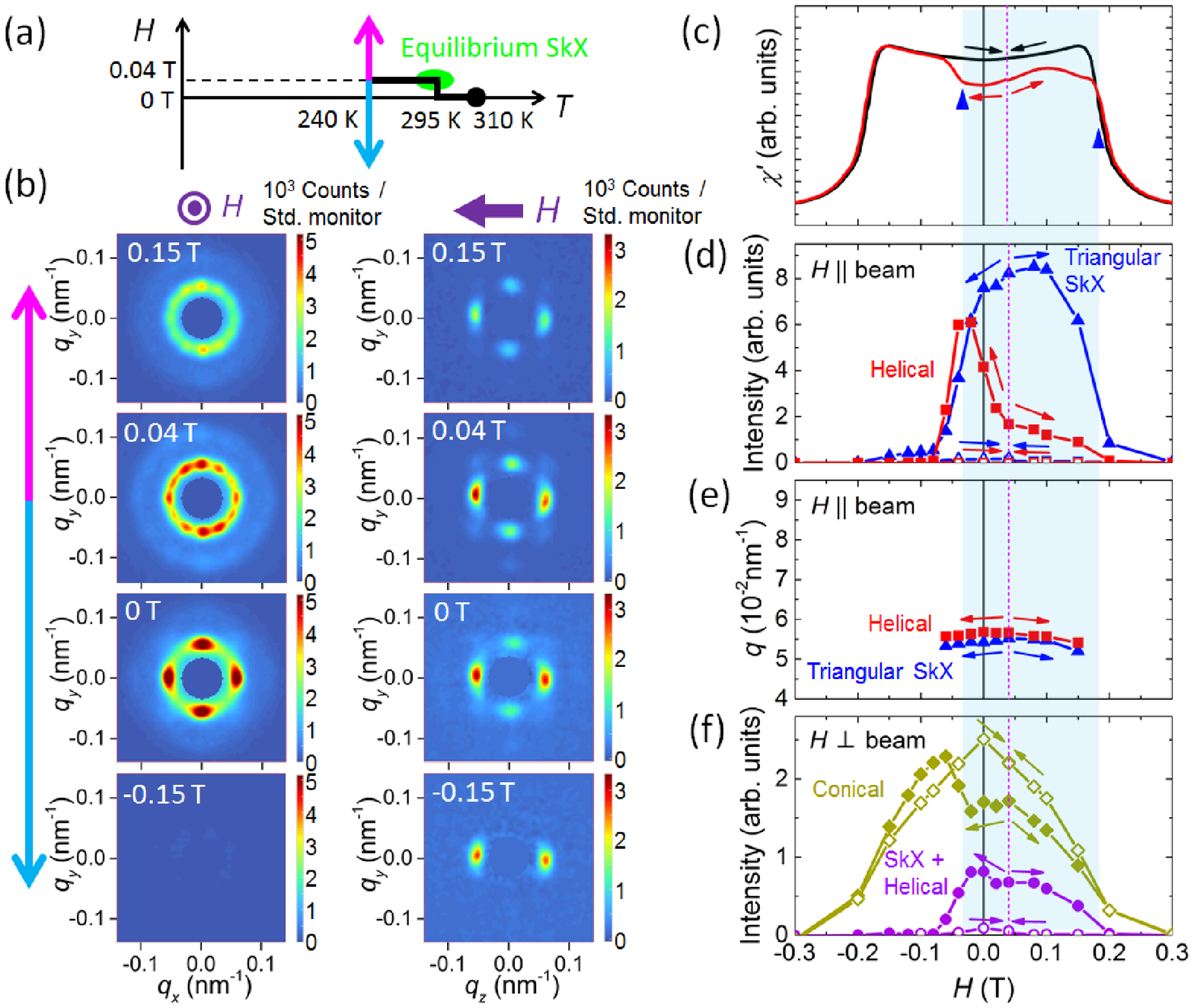}
\end{center}
\end{figure}
\textbf{Figure S5: Field dependence of the metastable SkX at 240 K after FC.} 
(a) Schematic illustration of the measurement process. Field-sweepings from 0.04 T to positive and negative field directions are denoted by pink and light blue arrows, respectively.
(b) SANS images obtained in the $H$ $\parallel$ beam and $H$ $\perp$ beam geometries and at selected fields. 
The full scale of the color plot is fixed for each measurement geometry.
(c) Field dependence of AC magnetic susceptibility. 
Field-sweepings from 0.04 T to $\pm$0.3 T (from $\pm$0.3 T to 0.04 T) are shown by red (black) lines. 
The blue triangles show the boundaries of the metastable triangular-lattice SkX state, which are plotted, together with the similar data points at different temperatures above 150 K, in the state diagram in Fig. 1(c). 
(d) Field dependence of the total scattered SANS intensity (defined similarly as in Fig. S2(c) and (d)) from the triangular-lattice SkX component (blue triangular symbols) and the helical component (red square symbols) in the $H$ $\parallel$ beam geometry. Field-sweepings from 0.04 T to $\pm$0.3 T (from $\pm$0.3 T to 0.04 T) are shown by closed (open) symbols. 
(e) Field dependence of $q$ (defined similarly as in Fig. S2(c) and S2(d)) of the triangular-lattice SkX component (blue triangular symbols) and the helical component (red square symbols). 
(f) Field dependence of the total scattered SANS intensity sum of the triangular-lattice SkX and helical components (purple circle symbols) appearing on the vertical axis, and the conical component (yellow diamond symbols) appearing on the horizontal axis, in the $H$ $\perp$ beam geometry. Field-sweepings from 0.04 T to $\pm$0.3 T (from $\pm$0.3 T to 0.04 T) are shown by closed (open) symbols. 

\newpage

\begin{figure}[tbp]
\begin{center}
\includegraphics[width=17cm]{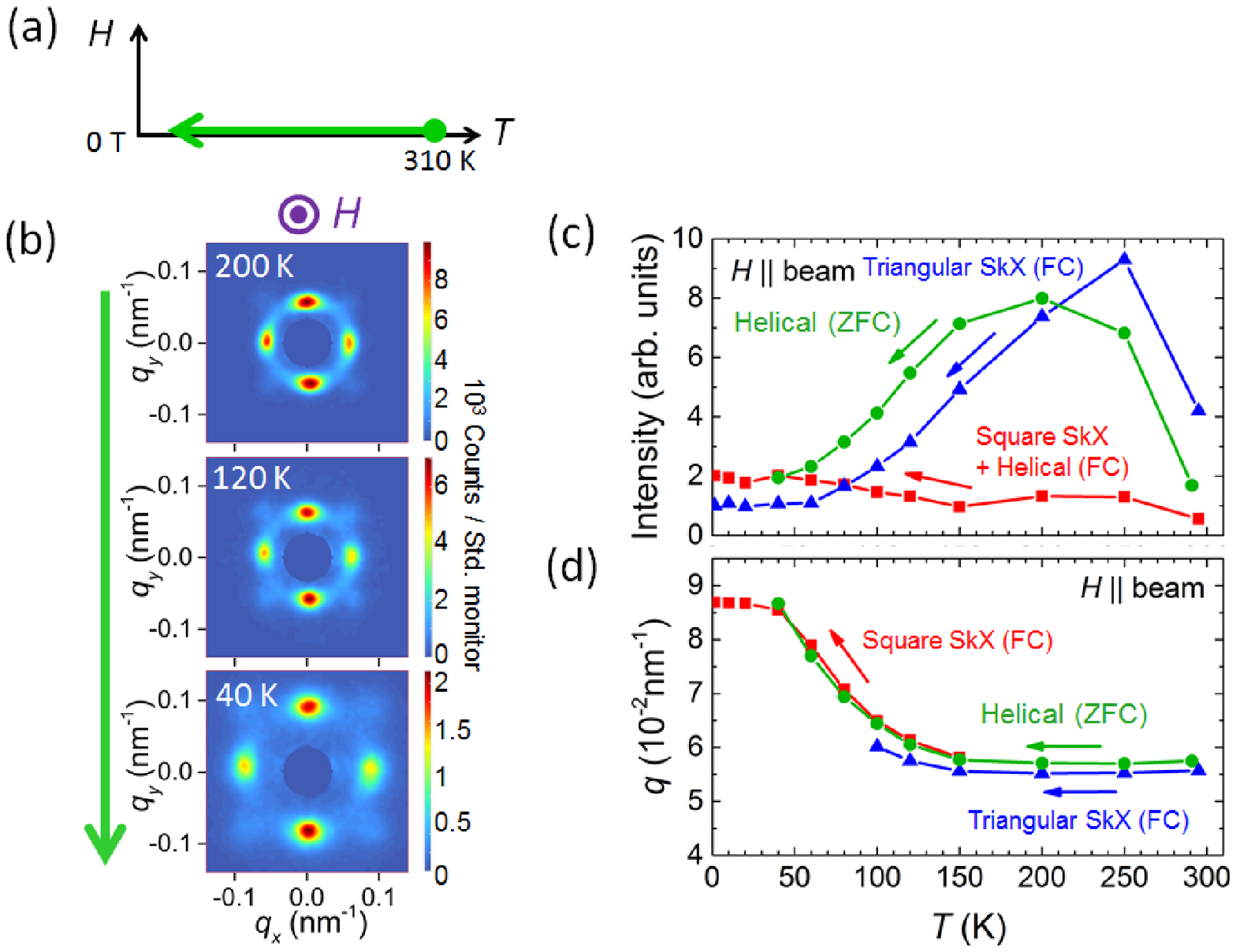}
\end{center}
\end{figure}
\textbf{Figure S6: SANS results in the ZFC process.} 
(a) Schematic illustration of the measurement process. 
(b) SANS images obtained in the $H$ $\parallel$ beam geometry and at selected temperatures. 
The full scale of the color plot varies between panels.
In this measurement, the rocking-angle was fixed at $\omega$ = 0$^\circ$ and and temperature was swept continuously. 
(c) Temperature dependence of the SANS intensity from the helical state (green circle symbols) in the $H$ $\parallel$ beam geometry (defined similarly as Fig. S2(d)). 
(d) Temperature dependence of $q$ of the helical state (green circle symbols) (defined similarly as Fig. S2(d)). 
In panels (c) and (d), the data of the FC process (Figs. 3(d) and (e)) are also plotted for comparison.

\newpage

\begin{figure}[tbp]
\begin{center}
\includegraphics[width=17cm]{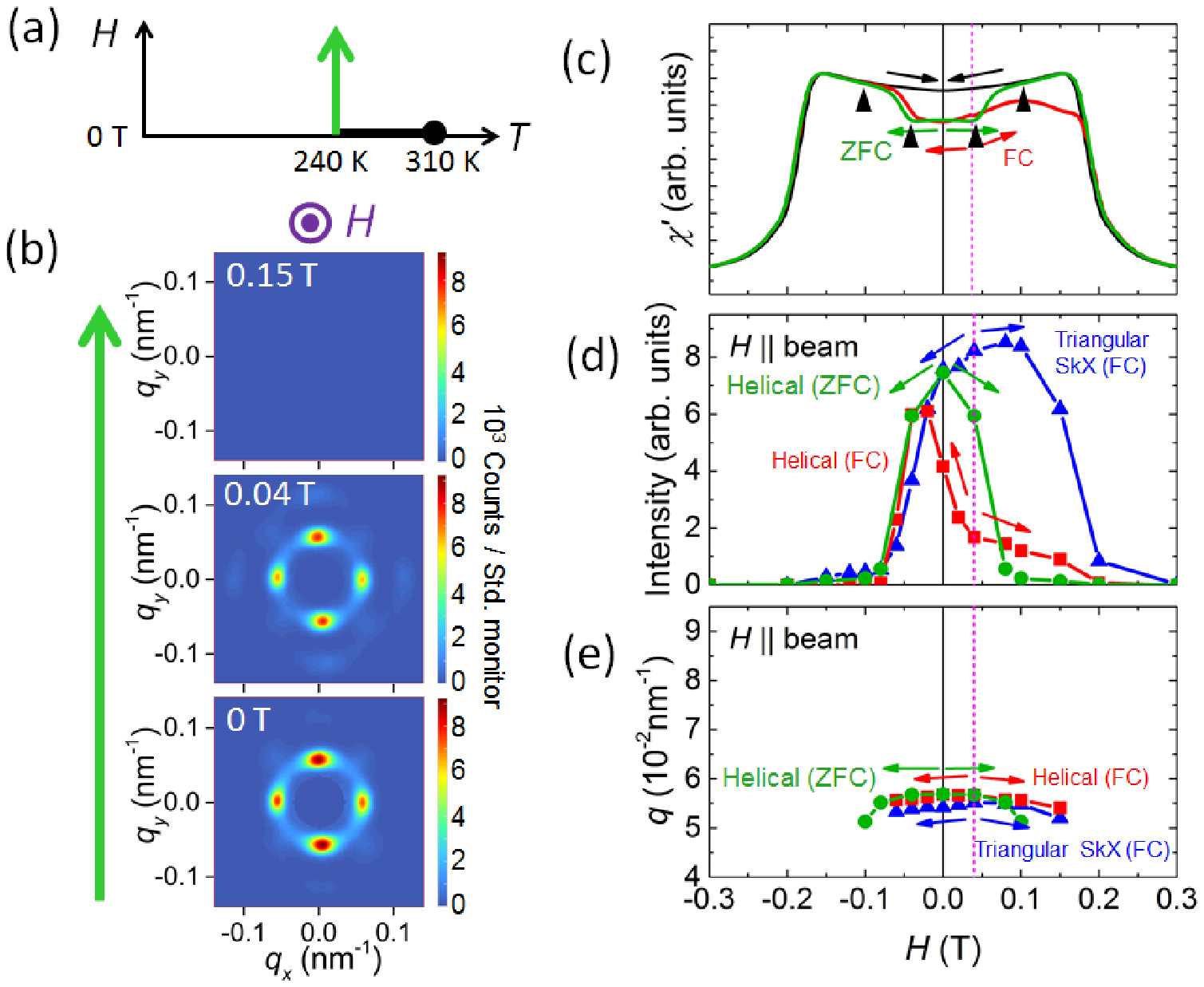}
\end{center}
\end{figure}
\textbf{Figure S7: SANS and AC magnetic susceptibility results in the field-sweeping process at 240 K after ZFC.} 
(a) Schematic illustration of the measurement process.
(b) SANS images obtained in the $H$ $\parallel$ beam geometry and at selected fields. 
The full scale of the color plot is fixed.
(c) Field dependence of AC magnetic susceptibility. 
Field-sweepings from 0 T to $\pm$0.3 T (from $\pm$0.3 T to 0 T) are shown by green (black) lines. 
The black triangles show the phase boundaries as plotted in the phase diagram in Fig. 1(b). 
(d) Field dependence of the total scattered SANS intensity observed from the helical state (green circle symbols) in the $H$ $\parallel$ beam geometry (defined similarly as Fig. S2(d)). 
(e) Field dependence of $q$ of the helical state (green circle symbols) (defined similarly as Fig. S2(d)). 
In panels (c), (d), and (e), the data obtained also at 240 K but after the FC process (Figs. S5(c), (d), and (e)) are also plotted for comparison.

\newpage

\begin{figure}[tbp]
\begin{center}
\includegraphics[width=17cm]{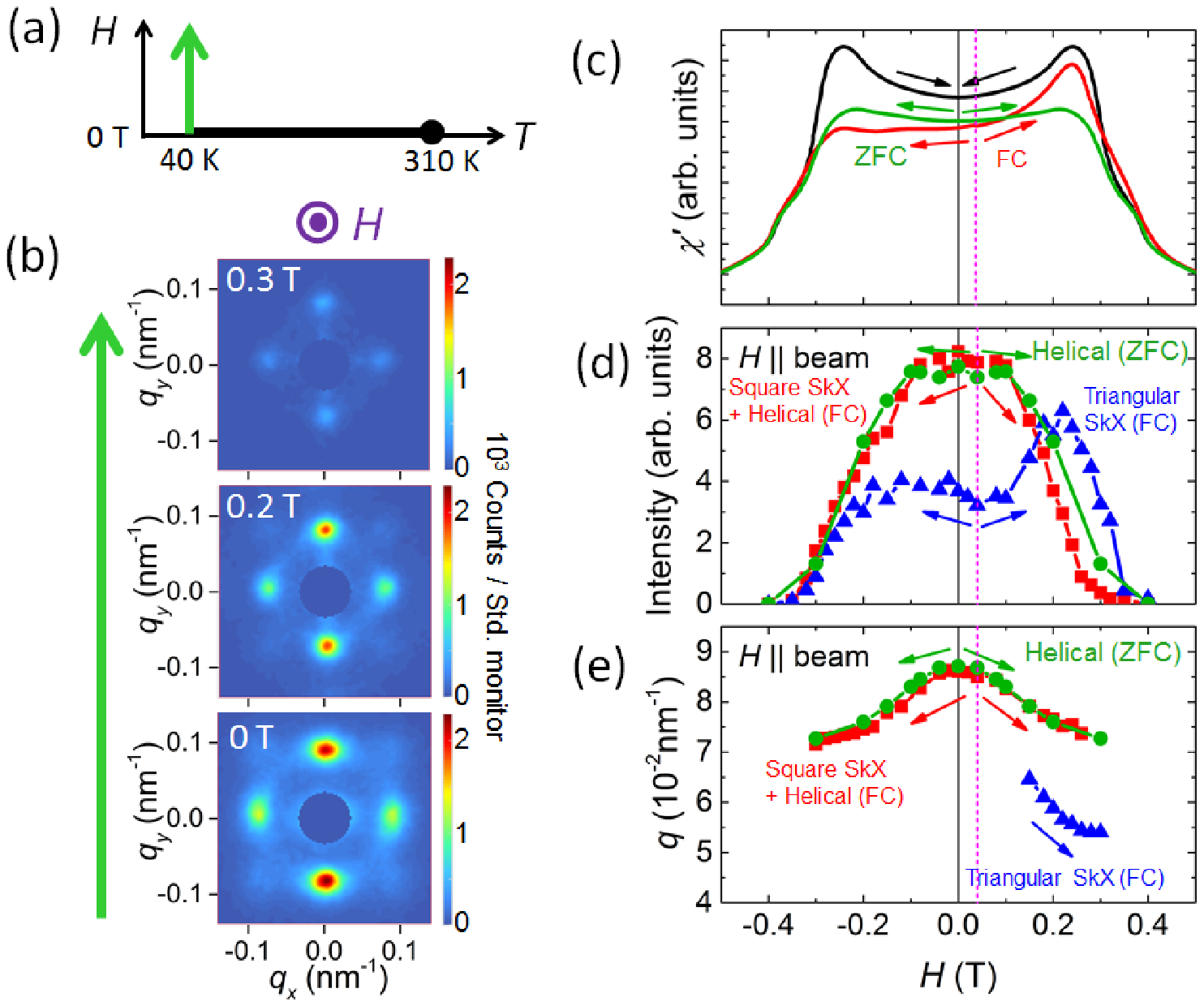}
\end{center}
\end{figure}
\textbf{Figure S8: SANS and AC magnetic susceptibility results in the field-sweeping process at 40 K after ZFC.} 
(a) Schematic illustration of the measurement process.
(b) SANS images obtained in the $H$ $\parallel$ beam geometry and at selected fields. 
The full scale of the color plot is fixed. 
(c) Field dependence of AC magnetic susceptibility. 
Field-sweepings from 0 T to $\pm$0.5 T (from $\pm$0.5 T to 0 T) are shown by green (black) lines. 
(d) Field dependence of the total scattered SANS intensity observed from the helical state (green circle symbols) in the $H$ $\parallel$ beam geometry (defined similarly as Fig. S2(d)). 
(e) Field dependence of $q$ of the helical state (green circle symbols) (defined similarly as Fig. S2(d)). 
In panels (c), (d), and (e), the data obtained also at 40 K but after the FC process (Figs. 4(c), (d), and (e)) are also plotted for comparison.

\newpage